\documentclass[showpacs,preprintnumbers,amsmath,amssymb,twocolumn,superscriptaddress,prb]{revtex4}
\usepackage{times}
\usepackage{graphicx}
\usepackage{amsfonts}
\usepackage{amsmath, amsthm, amssymb}
\usepackage{dsfont}
\usepackage{subfigure}

\newcommand{\vp}{\mathbf{p}}

\newcommand{\e}[1]{\mathrm{e}^{#1}}

\newcommand{\etal}{\emph{et al.}}

\def\i{\mathrm{i}}

\begin{document}
\title[Crossed Andreev reflection in superconducting graphene spin-valves: Spin-switch effect]{Crossed Andreev reflection in superconducting graphene spin-valves: Spin-switch effect}
\author{Jacob Linder}
\affiliation{Department of Physics, Norwegian University of
Science and Technology, N-7491 Trondheim, Norway}
\author{Malek Zareyan}
\affiliation{Institute for Advanced Studies in Basic Sciences, 45195-1159, Zanjan, Iran}
\author{Asle Sudb{\o}}
\affiliation{Department of Physics, Norwegian University of
Science and Technology, N-7491 Trondheim, Norway}

\date{Received \today}
\begin{abstract}
We consider the non-local quantum transport properties of a graphene superconducting spin-valve.
It is shown that one may create a spin-switch effect between perfect elastic co-tunneling (CT) and
perfect crossed Andreev-reflection (CAR) for all bias voltages in the low-energy regime by reversing the magnetization
direction in one of the ferromagnetic layers. This opportunity arises due the possibility of tuning the local
Fermi-level in graphene to values equivalent to a weak, magnetic exchange
splitting, thus reducing the Fermi surface for minority spins to a single point and rendering graphene to be half-metallic. Such an effect is not attainable in a conventional metallic
spin-valve setup, where the contributions from CT and CAR tend to cancel each other and noise-measurements
are necessary to distinguish these processes.

\end{abstract}

\maketitle
\section{Introduction}

Quantum entanglement \cite{rmp_entanglement} describes a scenario where the quantum states of two
objects separated in space are strongly correlated. These correlations can be exploited in emerging
technologies such as quantum computing, should one be able to spatially separate the entangled objects
without destroying the correlations. In a broader context, quantum entanglement could prove to be of practical importance in the fields of spintronics \cite{zutic_rmp_04} and information cryptography \cite{galindo_rmp_02}. It also holds a considerable interest from a purely fundamental physics point of view, prompting some of the more philosophically inclined discussions related to quantum theory and causality.
\par
Superconductors have been proposed as natural sources for entangled electrons
\cite{burkard_prb_00, recher_prb_01}, as Cooper pairs consist of two electrons that are both spin and
momentum-entangled. The Cooper pair can be spatially deformed by means of the crossed Andreev reflection
(CAR) process in superconducting heterostructures. In this scenario, an electron and hole excitation are 
two separate metallic leads are coupled by means of Andreev scattering processes at two spatially distinct
interfaces. Unfortunately, the signatures of CAR are often completely masked by a competing process known as
elastic co-tunneling (CT) which occur in the same type of heterostructures. In fact, the conductances stemming
from CT and CAR may cancel each other completely \cite{falci_epl_01}, thus necessitating the usage of
noise-measurements to find fingerprints of the CAR process in such superconducting heterostructures.
\par
Recently, graphene \cite{novoselov_science_04} has been studied as a possible arena for CAR-processes. In Ref. \cite{cayssol_prl_08}, it was shown how a three-terminal graphene sheet
containing $n$-doped, $p$-doped, and superconducting regions could be constructed to produce perfect CAR
for one particular resonant bias voltage. Also, the signatures of the CAR process in the noise-correlations
of a similar device were studied in Ref. \cite{benjamin_prb_08}. However, the role played by the spin degree
of freedom in graphene devices probing non-local transport has not been addressed so far. This is a crucial
point since it might be possible to manipulate the spin-properties of the system to interact with the spin-singlet
symmetry of the Cooper pair in a fashion favoring CAR.
\begin{figure}[t!]
\begin{center}
\scalebox{0.45}{
\includegraphics[width=19.0cm,clip]{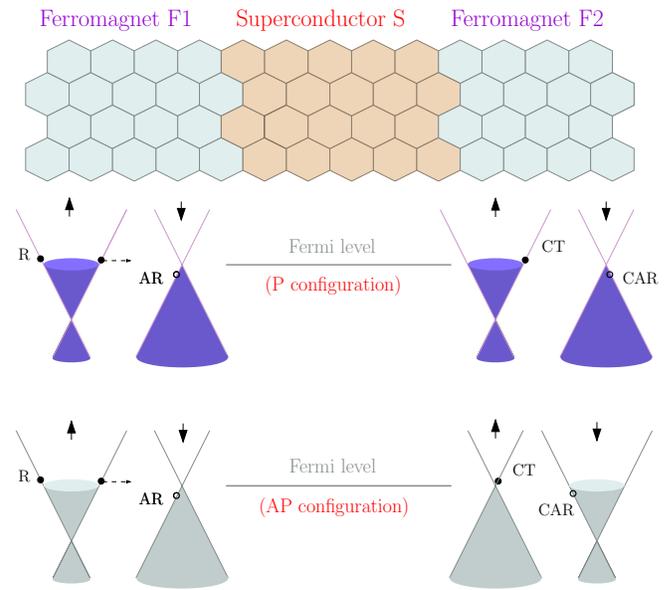}}
\end{center}
\caption{(color online) Proposed experimental setup for the spin switch effect between crossed Andreev reflection and elastic co-tunneling. Ferromagnetism and superconductivity are induced by the proximity effect to a host material. The induced exchange fields in the non-superconducting graphene regions are oriented either parallel or antiparallel with respect to each other. In the parallell alignment, the density of states vanishes for both normal Andreev reflection and crossed Andreev reflection processes, such that only elastic-cotunneling contributes to non-local transport. In the anti-parallel alignment, the density of states vanishes for both normal Andreev reflection and elastic co-tunneling, leaving only crossed Andreev reflection as the non-local transport channel.}
\label{fig:model}
\end{figure}
\par
In this paper, we show that precisely such an opportunity exists
-- it is possible to obtain a spin-switch effect between virtually
perfect CAR and perfect CT in a superconducting graphene spin
valve. In contrast to Ref. \cite{cayssol_prl_08}, this effect is
seen for all bias voltages in the low-energy regime rather than
just at one particular applied voltage difference. The key
observation is that the possibility of tuning the local
Fermi-level to values equivalent to a weak, magnetic exchange
splitting in graphene renders both the usual Andreev reflection
process and CT impossible. In contrast, this opportunity does not
exist in conventional conductors where the Fermi energy is large
and of order $\mathcal{O}$(eV). We show that graphene spin valves
provide a possibility for a unique combination of non-local
Andreev reflection and spin-dependent Klein tunneling
\cite{katsnelson_naturephys_06}. Our model is
shown in Fig. \ref{fig:model}, where ferromagnetism and
superconductivity are assumed to be induced by means of the
proximity effect \cite{tombros_nature_07,heersche_nature_06} to leads with the
desired properties. A similar setup was considered in Ref. \cite{bai_apl_08}, where the magnetoresistance of the system was studied.
\par
We organize this work as follows. In Sec. \ref{sec:theory}, we establish the theoretical framework which will be used to obtain the results. In Sec. \ref{sec:results}, we present our main findings for the non-local conductance in the graphene superconducting spin-valve with a belonging discussion of them. Finally, we summarize in Sec. \ref{sec:summary}.

\section{Theory}\label{sec:theory}
We consider a ballistic, two-dimensional graphene structure as shown in Fig. \ref{fig:model}.
In the left ferromagnetic region $x<0$, the exchange field is $\mathbf{h}=h_0\mathbf{z}$, while
it is $\mathbf{h}=\pm h_0\mathbf{z}$ in the right ferromagnetic region $x>L$. In the superconducting
region $0<x<L$, the order parameter is taken to be constant with a real gauge $\Delta=\Delta_0$. To
proceed analytically, we make the usual approximation of a step-function behavior at the interfaces
for all energy scales, i.e. the chemical potentials $\{\mu_F,\mu_S\}$, the exchange field $h_0$,
 and superconducting gap $\Delta_0$. This assumption is expected to be good when there is a substantial
 Fermi-vector mismatch between the F and S regions, as in the present case. To make contact with the
 experimentally relevant situation, we assume a heavily doped S region satisfying $\mu_S\gg\mu_F$.
\par

We use the Dirac-Bogoliubov de Gennes equations first employed in
Ref. \cite{beenakker_prl_06}. For quasiparticles with spin
$\sigma$, one obtains in an F$\mid$S graphene junction:
\cite{linder_prl_08, moghaddam_prb_08,zareyan_prb_08, asano_prb_08, zhang_prl_08}
\begin{align}\label{eq:bdg}
&\begin{pmatrix}
\hat{H}_\sigma(x)  & \sigma\Delta(x)\hat{1} \\
\sigma\Delta^*(x)\hat{1} & -\hat{H}_{-\sigma}(x) \\
\end{pmatrix}
\begin{pmatrix}
u^\sigma\\
v^{-\sigma}\\
\end{pmatrix} = \varepsilon
\begin{pmatrix}
u^\sigma\\
v^{-\sigma}\\
\end{pmatrix},
\end{align}
where 
\begin{align}
\hat{H}_\sigma(x) = v_\text{F}\mathbf{p} \cdot \hat{\boldsymbol{\sigma}} - [\mu(x)+\sigma h(x)] \hat{1}
\end{align}
 and $\hat{\ldots}$ denotes a $2\times2$ matrix. Here,
we have made use of the valley degeneracy and $\mathbf{p}$ is the
momentum vector in the graphene plane while
$\underline{\boldsymbol{\sigma}}$ is the
vector of Pauli matrices in the pseudospin space representing the
two A, B sublattices of graphene hexagonal structure. The
superconducting order parameter $\Delta(x)$ couples electron- and
hole-excitations in the two valleys ($\mp$)
located at the two inequivalent corners of the hexagonal Brillouin
zone. The $u^\sigma$ spinor describes the electron-like part of
the total wavefunction 
\begin{align}
\psi^\sigma = (u^\sigma,
v^{-\sigma})^\text{T},
\end{align}
 and in this case reads 
 \begin{align}
 u^\sigma =
(\psi_{A,+}^\sigma, \psi_{B,+}^\sigma)^\text{T}
\end{align}
 while
$v^{-\sigma} = \mathcal{T}u^\sigma$. Here, $^\text{T}$ denotes the
transpose while $\mathcal{T}$ is the time-reversal operator.
\par
From Eq. (\ref{eq:bdg}), one may now construct the quasiparticle
wavefunctions that participate in the scattering processes \cite{linder_prb}. We
consider positive excitation energies $\varepsilon\geq0$ with
incoming electrons of $n$-type, i.e. from the conduction band
$\varepsilon=v_\text{F}|\vp|-\mu_F$ (we set $v_\text{F}=1$ from now on). The
incoming electron from the left ferromagnet may either be
reflected normally or Andreev-reflection (AR). In the latter
process, it tunnels into the superconductor with another electron
situated at $(-\varepsilon)$, leaving behind a hole excitation
with energy $\varepsilon$. The scattering coefficients for these
two processes are $r_e$ and $r_h$, respectively, and the total
wavefunction may thus be written as:
\begin{align}
\psi_L = 
\begin{pmatrix}
1\\
\e{\i\theta}\\
0\\
0\\
\end{pmatrix}&\e{\i p_e^\sigma \cos\theta x} + 
r_e\begin{pmatrix}
1\\
-\e{-\i\theta}\\
0\\
0\\
\end{pmatrix}
\e{-\i p_e^\sigma \cos\theta x}\notag\\
&+r_h\begin{pmatrix}
0\\
0\\
1\\
\e{-\i\theta_A^\sigma}\\
\end{pmatrix}
\e{-\i p_h^\sigma \cos\theta_A^\sigma x},
\end{align}
where we have defined the wavevectors
\begin{align}
p_e^\sigma = \varepsilon + \mu_F + \sigma h_0,\; p_h^\sigma = \varepsilon-\mu_F+\sigma h_0.
\end{align}
We have omitted a common factor $\e{\i p_y y}$ for all wavefunctions. Similarly, assuming that the charge carriers in the right ferromagnetic region are also of the $n$-type, we obtain:
\begin{align}
\psi_R &= t_e
\begin{pmatrix}
1\\
\e{\i\theta}\\
0\\
0\\
\end{pmatrix}\e{\i p_e^{\pm\sigma} \cos\theta_N^{\pm\sigma}x} \notag\\
&+ t_h\begin{pmatrix}
0\\
0\\
1\\
\e{-\i\theta_A^{\pm\sigma}}\\
\end{pmatrix}\e{-\i p_h^{\pm\sigma} \cos\theta_A^{\pm\sigma}x}.
\end{align}
It should be noted that the AR hole is generated in the conduction band if $\varepsilon-\mu_F-\sigma h_0>0$ (retro-AR),
whereas it is generated in the valence band otherwise (specular-AR). The $\pm$ sign above refers to
parallell/antiparallell (P/AP) magnetization configuration.
\par
We assume that the superconducting region is heavily doped, $\mu_S \gg \mu_F +h_0$, which causes the
propagating quasiparticles to travel along the $x$-axis since the scattering angle in the superconductor
satisfies $\theta_S\to0$. We obtain the following wavefunction ($\lambda=\pm1$):
\begin{align}
\Psi_S &= \sum_{\lambda,\pm} l_\lambda^\pm 
\begin{pmatrix}
\e{\i\lambda\beta}\\
 \pm \e{\i\lambda\beta}\\
  1\\
  \pm 1\\
  \end{pmatrix}
  \e{\pm(\i \mu_S - \lambda\kappa)x},
\end{align}
where $\kappa =\sqrt{\Delta_0^2-\varepsilon^2}$ while 
\begin{align}
\beta = \text{acos}(\varepsilon/\Delta_0)
\end{align}
for subgap
energies $|\varepsilon|<\Delta_0$ and 
\begin{align}
\beta =-\i\text{acosh}(\varepsilon/\Delta_0)
\end{align}
 for supergap energies
$|\varepsilon|>\Delta_0$ .
\par
It is important to consider carefully the scattering angles in the problem. Since we assume translational
invariance in the $y$-direction, the $y$-component of the momentum is conserved. This gives
us 
\begin{align}
p_e^\sigma \sin\theta = p_h^\sigma \sin\theta_A^\sigma = p_e^{\pm\sigma}\sin\theta_N^{\pm\sigma}.
\end{align}
It is clear that the angle of transmission for the electrons in the right ferromagnet is equal to the
angle of incidence when the magnetizations are P, i.e. $\theta_N^\sigma = \theta$. Also, one infers that
there exists a critical angle above which the scattered waves become evanescent, i.e. decaying exponentially. This may be seen by observing that the scattering angles exceed $\pi/2$ (thus becoming imaginary) above a certain angle of incidence $\theta$.
For instance, the AR wave in the left ferromagnetic region becomes evanescent for angles of incidence
$\theta>\theta_\text{AR}^\sigma$, where the critical angle $\theta=\theta_\text{AR}^\sigma$ is obtained by setting $\theta_A^\sigma=\pi/2$ in the equation 
\begin{align}
p_e^\sigma \sin\theta = p_h^\sigma \sin\theta_A^\sigma,
\end{align}
 expressing conservation of momentum perpendicular to the interface. One finds that:
\begin{align}\label{eq:criticalAR}
\theta_\text{AR}^\sigma \equiv  |\text{asin}[(\varepsilon - \mu_F + \sigma h_0)/(\varepsilon+\mu_F+\sigma h_0)]|.
\end{align}
Thus, AR waves in the regime $\theta> |\theta_c^\sigma|$ do not contribute to any transport of charge. A similar
argument can be made for the transmitted electron wave-function in the right ferromagnetic region, corresponding
to the CT process, where the critical angle for this process becomes
\begin{align}
\theta_\text{CT}^\sigma \equiv |\text{asin}[(\varepsilon+\mu_F \pm \sigma h_0)/(\varepsilon + \mu_F +\sigma h_0)]|.
\end{align}
In the P configuration, the CT process thus always contributes to the transport of charge. Finally, the contribution
to transport of charge from CAR comes from the hole-wave function in the right ferromagnetic region, which becomes
evanescent for angles of incidence above the critical angle
\begin{align}
\theta_\text{CAR}^\sigma > |\text{asin}[(\varepsilon - \mu_F \pm \sigma h_0)/(\varepsilon+\mu_F+\sigma h_0)]|.
\end{align}
In the P configuration, this criteria is the same as the vanishing of local AR expressed by Eq. (\ref{eq:criticalAR}).

\section{Results and Discussion}\label{sec:results}

Intuitively, one might expect that the most interesting phenomena occur when the exchange field $h_0$ is comparable
in magnitude to the chemical potential $\mu_F$. If $\mu_F \gg h_0$, the effect of the exchange field should be minor
and the AR is never specular. In contrast, the situation becomes quite fascinating when we consider the case
$\mu_F=h_0$ under the assumption of a doped situation $\mu_F\gg(\varepsilon,\Delta_0)$. First of all, the incoming
quasiparticles from the left ferromagnetic region are completely dominated by the majority spin carriers
$\sigma=\uparrow$, since the density of states (DOS) for $\sigma=\downarrow$ electrons vanishes at the Fermi
level. Since $\mu_F=h_0$, the AR process is suppressed for all incoming waves as $\theta_\text{AR}^\uparrow\to 0$.
We now show how the fate of the cross-conductance in the right ferromagnetic region depends crucially on whether
the magnetization configuration is P or AP. In the P configuration, we see that $\theta_\text{CAR}^\uparrow\to0$,
which means that the transport is purely governed by the CT process. In the AP configuration, we see
that $\theta_\text{CT}^\uparrow \to0$, which means that the transport is mediated purely by the CAR process.
This suggests a remarkable spin-switch effect -- by reversing the direction of the field in the right ferromagnet,
one obtains an abrupt change from pure CT to pure CAR processes mediating the transport of charge. In each case,
there is no local AR in the left ferromagnetic region. In the standard metallic case, the distinct signatures for
the CT and CAR contributions are masked by each other, and it becomes necessary to resort to noise-measurements in
order to say something about the contribution from each process. In the present scenario, we have showed how it is
possible to separate the two contributions directly by a simple spin-switch effect which is commonly employed in
experimental work on F$\mid$S heterostructures.

\begin{figure}[t!]
\begin{centering}\includegraphics[width=8cm]{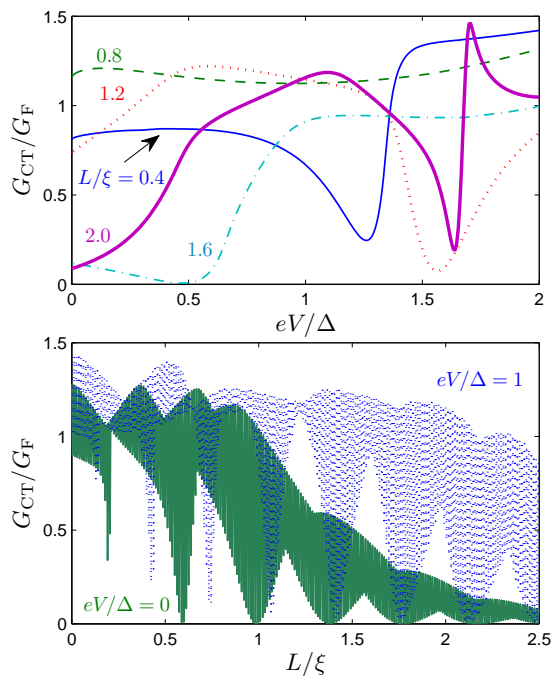}\par\end{centering}
\caption{\label{fig:P}{\small (color online) Plot of the conductance for CT processes $G_{\text{CT}}/G_{\text{F}}$ versus bias voltage in the upper panel and versus length of the S region in the lower panel. Here, we consider the P alignment and $\mu_F=h_0$ such that $G_{\text{CAR}}\to0$.
 }}\label{fig:P}
\end{figure}
\par
\begin{figure}[t!]
\begin{centering}\includegraphics[width=8cm]{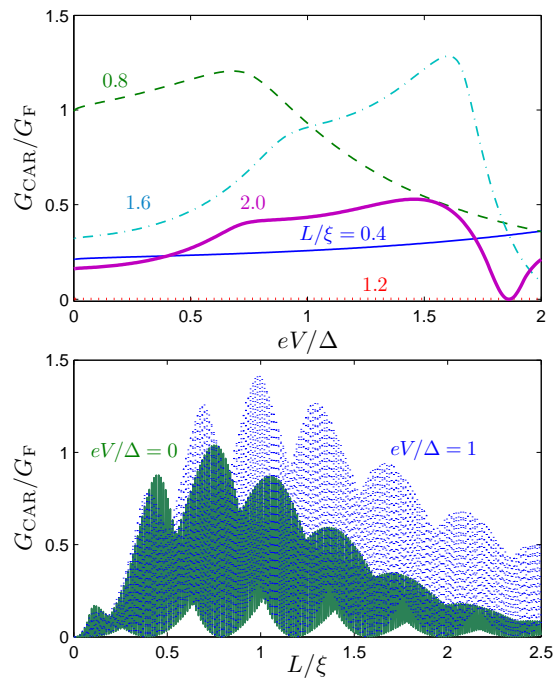}\par\end{centering}
\caption{\label{fig:resonance}{\small (color online) Plot of the conductance for CAR processes
$G_{\text{CAR}}/G_{\text{F}}$ versus bias voltage in the upper panel and versus length of the S
region in the lower panel. Here, we consider the AP alignment and $\mu_F=h_0$ such that $G_{\text{CT}}\to0$.
}}\label{fig:AP}
\end{figure}

Let us now evaluate the conductance in the P and AP configuration quantitatively by using
\begin{align}\label{eq:conductance}
G_\text{CAR}/G_\text{F} &= \sum_{\sigma} (G^{\sigma}/G_\text{F})
\int_{-\pi/2}^{\pi/2}{d\theta }\cos\theta |t_h|^2,
\end{align}
where we have introduced
\begin{align}
G^{\sigma}=e^2N^{\sigma}(eV)/\pi
\end{align}
as the spin-$\sigma$
normal-state conductance that takes into account the valley
degeneracy, in addition to
\begin{align}
G_{\rm F}=G^{+}+G^{-}.
\end{align}
 The density of states is
determined by 
\begin{align}
N^{\sigma}(\varepsilon)=|\varepsilon+\mu_F+\sigma h|W/(\pi  v_{\rm F}),
\end{align}
where $W$ is the width of the
junction. The expression for $G_{\text{CT}}$ is obtained by replacing $t_h$ with $t_e$ in Eq. (\ref{eq:conductance}). Since we here consider the case $\mu_F=h_0$ and $h_0\gg (\varepsilon,\Delta_0)$, the formulas for the $G_{\text{CAR}}$ and $G_{\text{CT}}$ may be simplified since $G_-\ll G_+$. Also, since the DOS vanishes for minority spins for the injected electrons, only $\sigma=\uparrow$ contributes for incoming electrons. The crucial point here is that in the P alignment, $G_{\text{CAR}} \to 0$ and $G_{\text{CT}}\neq0$ such that 
\begin{align}
|r_e|^2 + |t_e|^2 = 1,
\end{align} while in the AP alignment $G_{\text{CAR}} \neq 0$ and $G_{\text{CT}} \to 0$ such that 
\begin{align}
|r_e|^2 + |t_h|^2 = 1.
\end{align}
In the actual numerical calculations, we use $h_0/\Delta_0=50$ and $\mu_S/\Delta_0=500$. Assuming a value of $\Delta_0=0.1$ meV for the proximity-induced gap, this corresponds to an exchange splitting of $h_0=5$ meV in the F regions and a doping level $\mu_S=50$ meV in the S region, which should be experimentally feasible \cite{haugen_prb_08} and well within the range of the validity for the linear dispersion relation in graphene. In Fig. \ref{fig:P}, we plot the cross-conductance $G_{\text{CT}}/G_\text{F}$ in the P alignment both as a function of bias voltage and width of the S region. The same thing is done for $G_{\text{CAR}}/G_\text{F}$ in the AP alignment in Fig. \ref{fig:AP}. In both cases, the magnitude of the conductance varies strongly when considering different widths $L$ due to the fast oscillations which pertain to the formation of resonant transmission levels inside the superconductor. Also, it is seen that while the CT process is favored for short junctions $L/\xi\ll1$, the CAR process is suppressed in this regime in favor of normal reflection. Upon increasing the junction width, the CT conductance drops while the CAR conductance peaks at widths $L\sim\xi$.
The remarkable aspect is that it is possible to switch between these two scenarios of exclusive CT and exclusive CAR simply by reversing the direction of magnetization in one of the ferromagnetic layers.
\par
In order to obtain analytical results, we have assumed that the
Coulomb interaction and charge inhomogeneities may be neglected.
It would be challenging to obtain a truly homogeneous chemical
potential in a graphene sheet, and electron-hole puddles appear to
be an intrinsic feature of graphene sheets \cite{puddles}.
Moreover, it has been speculated that such charge inhomogeneities
may play an important role with regard to limiting the transport
characteristics of graphene \cite{castro} near the Dirac points.
However, for our purposes this is actually beneficial -- it is
precisely the suppression of charge and spin 
transport at Fermi level for the Andreev reflection and
co-tunneling process which renders possible the spin-switch
effect. Therefore, we do not expect that the inclusion of charge
inhomogeneities should alter our results qualitatively. Finally, we note that since the spin of the charge-carriers in each of the non-superconducting graphene sheets are practically speaking fixed due to the vanishing DOS for minority spins, the spin-switch effect for CAR and EC predicted in this paper can not be directly related to entanglement. Nevertheless, it constitutes a clear non-local signal for quantum transport which can be probed experimentally, and should be helpful in identifying clear-signatures of the mesoscopic CAR phenomenon.

\section{Summary}\label{sec:summary}
To summarize, we have considered non-local quantum transport in a
graphene superconducting spin-valve. We have shown how one may
create a spin-switch effect between perfect elastic co-tunneling
and perfect crossed Andreev-reflection for all applied bias
voltages by reversing the magnetization direction in one of the
ferromagnetic layers. The basic mechanism behind this effect is that the local Fermi-level in graphene may be tuned so that the Fermi surface for minority spins reduces to a single point in the presence of a weak, magnetic exchange splitting.
This is very distinct from the equivalent
spin valve structures in conventional metallic systems, where
noise-measurements are required to clearly distinguish between
these processes.

\acknowledgments

 J.L. and A.S. were
supported by the Norwegian Research Council Grant Nos. 158518/431,
158547/431, (NANOMAT), and 167498/V30 (STORFORSK).

\end{document}